\PassOptionsToPackage{unicode}{hyperref}
\PassOptionsToPackage{hyphens}{url}
\PassOptionsToPackage{dvipsnames,svgnames,x11names}{xcolor}
\documentclass[
]{article}
\usepackage{amsmath,amssymb}
\usepackage{lmodern}
\usepackage{iftex}
\ifPDFTeX
  \usepackage[T1]{fontenc}
  \usepackage[utf8]{inputenc}
  \usepackage{textcomp} 
\else 
  \usepackage{unicode-math}
  \defaultfontfeatures{Scale=MatchLowercase}
  \defaultfontfeatures[\rmfamily]{Ligatures=TeX,Scale=1}
\fi
\IfFileExists{upquote.sty}{\usepackage{upquote}}{}
\IfFileExists{microtype.sty}{
  \usepackage[]{microtype}
  \UseMicrotypeSet[protrusion]{basicmath} 
}{}
\makeatletter
\@ifundefined{KOMAClassName}{
  \IfFileExists{parskip.sty}{%
    \usepackage{parskip}
  }{
    \setlength{\parindent}{0pt}
    \setlength{\parskip}{6pt plus 2pt minus 1pt}}
}{
  \KOMAoptions{parskip=half}}
\makeatother
\usepackage{xcolor}
\NewDocumentCommand\citeproctext{}{}
\NewDocumentCommand\citeproc{mm}{%
  \begingroup\def\citeproctext{#2}\cite{#1}\endgroup}
\makeatletter
 \let\@cite@ofmt\@firstofone
 \def\@biblabel#1{}
 \def\@cite#1#2{{#1\if@tempswa , #2\fi}}
\makeatother
\newlength{\cslhangindent}
\newlength{\csllabelwidth}
\newenvironment{CSLReferences}[2] 
 {\begin{list}{}{%
  \setlength{\itemindent}{0pt}
  \setlength{\leftmargin}{0pt}
  \setlength{\parsep}{0pt}
  \ifodd #1
   \setlength{\leftmargin}{\cslhangindent}
   \setlength{\itemindent}{-1\cslhangindent}
  \fi
  \setlength{\itemsep}{#2\baselineskip}}}
 {\end{list}}
\usepackage{calc}

\ifLuaTeX
\usepackage[bidi=basic]{babel}
\else
\usepackage[bidi=default]{babel}
\fi
\babelprovide[main,import]{american}

\def\languageshorthands#1{}
\ifLuaTeX
  \usepackage{selnolig}  
\fi
\IfFileExists{bookmark.sty}{\usepackage{bookmark}}{\usepackage{hyperref}}
\IfFileExists{xurl.sty}{\usepackage{xurl}}{} 
\hypersetup{
  pdftitle={easyCHEM: A Python package for calculating chemical
equilibrium abundances in exoplanet atmospheres},
  pdfauthor={Elise Lei, Paul Mollière},
  pdflang={en-US},
  colorlinks=true,
  linkcolor={Maroon},
  filecolor={Maroon},
  citecolor={Blue},
  urlcolor={Blue},
  pdfcreator={LaTeX via pandoc}}

\title{easyCHEM: A Python package for calculating chemical equilibrium
abundances in exoplanet atmospheres}

\definecolor{c53baa1}{RGB}{83,186,161}
\definecolor{c202826}{RGB}{32,40,38}

\usepackage[affil-it]{authblk}
\usepackage{orcidlink}
\author[1%
  *%
  ]{Elise Lei%
    }
\author[2%
  *%
  \ensuremath\mathparagraph]{Paul Mollière%
    \,\orcidlink{0000-0003-4096-7067}\,%
    }

\affil[1]{Mines Paris - PSL University, Paris, France%
  }
\affil[2]{Max Planck Institute for Astronomy, Heidelberg, Germany%
  }
\affil[$\mathparagraph$]{Corresponding author: %
}
\affil[*]{These authors contributed equally.}
\date{29 August 2024}

\begin{document}
\maketitle

\section{Summary}\label{summary}

For modeling the spectra of exoplanets one must know their atmospheric
composition. This is necessary because the abundance of molecules,
atoms, ions and condensates is needed to construct the total
cross-section for the interaction between electro-magnetic radiation and
matter. In addition, when solving for the temperature structure of an
atmosphere the so-called adiabatic temperature gradient must be known,
which prescribes the pressure-temperature dependence in convectively
unstable regions\footnote{The corresponding convective region is called
  ``troposphere'' on Earth.}. Depending on the planetary properties, the
composition and adiabatic gradients may be well described by equilibrium
chemistry, which means that chemical reactions occur faster than any
other processes in the atmosphere, such as mixing. What is more, the
equilibrium assumption often serves as a useful starting point for
non-equilibrium calculations. Efficient and easy-to-use codes for
determining equilibrium abundances are therefore needed.

\section{Statement of need}\label{statement-of-need}

easyCHEM\footnote{\url{https://easychem.readthedocs.io}} is a Python package for calculating chemical equilibrium
abundances (including condensation) and adiabatic gradients by
minimization of the so-called Gibbs free energy. easyCHEM implements the
equations presented in Gordon \& McBride
(\citeproc{ref-Gordon:1994}{1994}) (which details the theory behind
NASA's
\href{https://www1.grc.nasa.gov/research-and-engineering/ceaweb/}{CEA
equilibrium code}) from scratch in modern Fortran, and wraps them in
Python to provide an easy-to-use package to the community. For efficient
matrix inversion, required for the Gibbs minimization, easyCHEM
incorporates the optimized \texttt{dgesv} routine of the
\href{https://netlib.org/lapack/explore-html-3.6.1/d7/d3b/group__double_g_esolve_ga5ee879032a8365897c3ba91e3dc8d512.html}{LAPACK
library}. Users can interact with easyCHEM's \texttt{ExoAtmos} class to
calculate atmospheric compositions with just a few lines of code. Users
have full control over the atmospheric elemental composition and
chemical reactant selection.

The CEA code itself is written in a fixed-form FORTRAN77 style and
interacted with the user via input files. In Mollière et al.
(\citeproc{ref-Molliere:2017}{2017}) we introduced easyCHEM as a
from-scratch implementation using the Gordon \& McBride
(\citeproc{ref-Gordon:1994}{1994}) equations and modern Fortran, without
the need for input files for run specification. easyCHEM was benchmarked
with CEA, leading to identical results. It was also successfully
benchmarked with other equilibrium chemistry codes in Baudino et al.
(\citeproc{ref-Baudino:2017}{2017}). easyCHEM calculations have been
used in many publications, such as Nasedkin et al.
(\citeproc{ref-Nasedkin:2024}{2024}) and de Regt et al.
(\citeproc{ref-deRegt:2024}{2024}), to name a few recent ones. Here we
report on the Python-wrapped version and make all of its source code
public, to further increase its usefulness and accessibility.

We note that other open-source Python packages for computing chemical
equilibrium abundances exist, such as
\href{https://github.com/dzesmin/TEA}{TEA}
(\citeproc{ref-Blecic:2016}{Blecic et al., 2016}) or
\href{https://newstrangeworlds.github.io/FastChem/index.html}{FastChem}
(\citeproc{ref-Kitzmann:2024}{Kitzmann et al., 2024}).

\section*{References}\label{references}
\addcontentsline{toc}{section}{References}

\phantomsection\label{refs}

\end{document}